\documentclass[usenatbib,useAMS]{mn2e}

\usepackage{booktabs}
\usepackage{natbib}
\usepackage{amsmath}
\usepackage{amssymb}
\usepackage{graphicx}
\usepackage{subfigure}
\usepackage{color}
\usepackage{epsfig}
\bibpunct[,]{(}{)}{;}{a}{}{,}
\allowdisplaybreaks[1]

\title[Unstable magnetosonic waves in radiative envelopes]
{Sub-photospheric fluctuations in magnetized radiative envelopes: contribution from unstable magnetosonic waves}
\author[Sen, Fern\'andez, \& Socrates]{Koushik Sen$^{1,2}$, Rodrigo Fern\'andez$^{1}$\thanks{E-mail: rafernan@ualberta.ca}, 
and Aristotle Socrates$^{3}$\\
$^{1}$Department of Physics, University of Alberta, Edmonton, AB T6G 2E1, Canada.\\ 
$^{2}$Department of Physics, Indian Institute of Technology, Kharagpur, West Bengal, 721302, India.\\ 
$^{3}$Jackson, WY 83001, USA.}

\begin{document}

\date{To be submitted to MNRAS}

\pagerange{\pageref{firstpage}--\pageref{lastpage}} \pubyear{2017}

\maketitle

\label{firstpage}

\begin{abstract}
We examine the excitation of unstable magnetosonic waves in the radiative envelopes of 
intermediate- and high-mass stars with a magnetic field of $\sim $\,kG strength.
Wind clumping close to the star and microturbulence can often be accounted for
when including small-scale, sub-photospheric density or velocity perturbations. Compressional waves 
-- with wavelengths comparable to or shorter than the gas pressure scale height --
can be destabilized by the radiative flux in optically-thick media when a magnetic field is present, 
in a process called the Radiation-Driven Magneto-Acoustic Instability (RMI).
The instability does not require radiation or magnetic pressure to dominate over gas pressure,
and acts independently of sub-surface convection zones. 
Here we evaluate the conditions for the RMI to operate 
on a grid of stellar models covering a mass range $3-40M_\odot$ at solar metallicity.
For a uniform 1\,kG magnetic field, fast magnetosonic modes are unstable 
down to an optical depth of a few tens, while unstable slow modes extend beyond the depth of the iron convection zone.
The qualitative behavior is robust to magnetic field strength variations by a factor of a few.
When combining our findings with previous results for the saturation amplitude of the RMI, 
we predict velocity fluctuations in the range $\sim 0.1-10$\,km\,s$^{-1}$. These amplitudes
are a monotonically increasing function of the ratio of radiation to gas pressure, or alternatively, of the
zero-age main sequence mass. 
\end{abstract}



\section{Introduction}
\label{s:intro}

Our understanding of stellar magnetism and mass loss in massive stars
has evolved significantly during the last two decades. Spectropolarimetric
surveys have shown that $\sim 7\%$ of O-type stars host
persistent magnetic fields of $\sim$kG strength (e.g., \citealt{wade_2016,grunhut_2017}). 
Similarly, about 10\% of all A- and B-type stars are chemically peculiar and
host surface magnetic fields $\gtrsim 300$\,G 
(e.g., \citealt{landstreet_2007,auriere_2007,sikora_2017}).
In addition to large-scale, stable fields, rapid variability in O- to A-type stars 
suggests that smaller-scale fields and fluctuations, presumably originating
below the stellar surface, are present in a larger fraction of intermediate- and
high-mass stars (e.g., \citealt{rami_2014,sudnik_2016,sikora_2017b,rami_2018}).

Mass loss in massive stars has also been subject to revision, with significant
implications for massive star evolution (e.g., \citealt{smith_2014,renzo_2017}). The smooth
line-driven winds \citep{castor1975} that are often used in stellar evolution models are found
to exceed observational estimates by factors of a few when clumping effects are taken 
into account (e.g., \citealt{najarro_2011,cohen_2014}).
Clumping is expected from an intrinsic instability
of line-driven winds (the \emph{line deshadowing instability} or LDI; \citealt{lucy_1970,macgregor1979,owocki_1984}). 
Pervasive clumping is indeed found in time-dependent wind simulations that properly account for LDI effects
(e.g., \citealt{owocki1988,dessart_2005,sundqvist_2017}). However, agreement with observationally-inferred clumping
factors is only obtained, particularly at the base of the wind, when 
an additional source of perturbations is assumed at the photosphere \citep{sundqvist_2013}.

Sub-photospheric fluctuations are most often assumed to be the consequence of
wave excitation (primarily g-modes) by sub-surface convection zones at the iron opacity peak \citep{C09}.
The existence of this opacity peak is well established \citep{iglesias_1992,badnell_2005}, 
and the excitation of gravity waves at convective-radiative boundaries has been studied 
experimentally (e.g., \citealt{townsend_1966,lebars_2015}) and theoretically
(e.g., \citealt{goldreich_1990,lecoanet_2013}). The effect of these convection zones is
stronger with increasing luminosity and lower effective temperature \citep{C09}.
These convection zones could also be responsible for generating sub-surface magnetic
fields via dynamo action \citep {cantiello_2011} and thereby account for localized co-rotating 
magnetic structures. 

Convection zones at the iron opacity peak disappear below a metallicity-dependent luminosity, 
however, and the presence of strong large-scale fields could even inhibit
the development of these convection zones, since they are not very efficient while in the main 
sequence \citep{cantiello_2011}. Small-scale destabilization
of purely acoustic waves is still possible for sufficiently strong 
radiative driving, if the opacity has the right density dependence 
(\citealt[ hereafter BS03]{BS03}; \citealt{suarez_2013}). 

BS03 found that the presence of a magnetic field 
enables an additional source of small-scale ($\lesssim$ gas pressure scale height), 
sub-photospheric fluctuations in stars with radiative envelopes. 
By performing a local linear stability analysis, BS03 identified the
physical mechanisms involved and showed that
the radiative flux can destabilize magnetosonic waves
when the radiative force performs work on the wave velocity
component along the background magnetic field. 
This instability had originally been identified in the context of highly-magnetized, 
radiation-dominated media \citep{PS73,arons_1992,klein_1996} and the form of its non-linear development 
led to it being called the \emph{photon bubble instability}. A number
of studies have focused on the photon bubble instability in environments
where radiation and magnetic pressure dominate over the gas pressure 
\citep{hsu1997,gammie_1998,BS01,begelman2001,davis2004,turner2005,turner2007,jiang2012}. 
The instability mechanism is quite general, however, and it is predicted to occur
even when the gas pressure dominates over both radiation and
magnetic pressure, as is the case in the radiative envelopes of intermediate- and high-mass
stars (BS03, \citealt{turner2004}).

\citet[ hereafter FS13]{FS13} studied the non-linear development of this instability over
a wide range of conditions using local two-dimensional, time-dependent radiation-magnetohydrodynamic
simulations in the diffusion regime. They confirmed the theoretical predictions of BS03 in that 
instability can occur even when the gas pressure is dominant. To better reflect
the driving mechanism at wavelengths comparable or smaller than the gas pressure scale-height,
they designated the process as the \emph{radiation-driven magneto-acoustic instability} (RMI), given that no
buoyancy is involved in this regime. The saturation amplitude of the RMI is a monotonic function 
of the ratio of radiation to gas pressure, and peaks when the magnetic pressure is comparable 
to the gas pressure. While FS13 pointed out the connection to sub-photospheric flucutations
in magnetized stellar envelopes, they did not apply their findings to realistic
stellar models.

Here we set out to explore the importance of the RMI in the radiative envelopes of magnetized
massive stars in which radiation pressure has a moderate influence. We focus on 
stars for which the radiative flux is not too strong to generate a significant density
inversion (which would significantly modify the background state; c.f. \citealt{jiang_2015, jiang_2017}), 
nor too weak that the saturation amplitude becomes irrelevant. These conditions are also 
less favorable for the development of a significant sub-surface iron convection zone. 
We evaluate the RMI instability conditions on a grid of stellar evolution models, and estimate the 
magnitude of the density and velocity fluctuations expected based on the time-dependent simulation 
results of FS13. The goal is to map out the regions of parameter space in which the RMI is likely to
play an important role in magnetized stellar envelopes.

The paper is organized as follows. Section 2 provides a brief overview of the RMI.
Section 3 describes the stellar models and physical assumptions used. 
Results are presented in Section 4, followed by a summary and discussion
in Section 5.

\section{Overview of the RMI}
\label{s:overview}

In magnetohydrodynamics (MHD), the dispersion relation for magnetosonic modes is 
\begin{equation}
\label{eq:magnetosonic_disprel}
\omega_0^2 = \frac{k^{2}}{2} \left[ (c_{\rm s}^2 + v_{\rm A}^2) \pm 
  \sqrt{(c_{\rm s}^2 + v_{\rm A}^2)^2 - 4c_{\rm s}^2(\hat{k}\cdot\mathbf{v_{\rm A}})^{2}} \right]
\end{equation}
where $\omega_0$ is the mode frequency, $k$ is the wave number and $\hat k$ the unit wave vector,
$c_{\rm s}$ is the sound speed, and $v_{\rm A} = B/\sqrt{4\pi\rho}$ is the Alfv\'en
speed, with $B$ the magnitude of the magnetic field and $\rho$ the fluid density.
Positive and negative signs in equation~(\ref{eq:magnetosonic_disprel}) define 
the \emph{fast} and \emph{slow} magnetosonic branches, respectively, given
the magnitude of the implied phase velocity $v_{\rm ph} = \omega_0/k$. 
Magnetosonic modes reduce to sound waves in the limit of vanishing magnetic field.

BS03 showed that in a stably-stratified and optically-thick medium, 
magnetosonic modes with wavelengths shorter than the gas pressure scale height 
can be destabilized by a background radiation field when radiative diffusion 
is rapid compared to the mode frequency. The instability mechanism is quite 
general, and relies on the coupling of the wave displacement vector
along the background  magnetic field and the perturbation to the 
radiative flux, which results in work done on the fluid oscillation.
Instability results when driving by the radiative flux overcomes
damping from radiative diffusion. The instability operates under a broad range of 
conditions, including weakly-magnetized media in which radiation pressure is sub-dominant, 
such as in the envelopes of massive stars. 

The condition that diffusion occurs more rapidly than the mode oscillation is
\begin{equation}
\label{eq:rapid_diffusion_condition}
\omega_{\rm diff} \equiv \dfrac{c k^{2}}{3\kappa_{\rm F}\rho} \gg \omega_0,
\end{equation}
where $\kappa_{\rm F}$ is the flux-mean opacity. This condition generally sets the maximum
optical depth at which the RMI can operate. At small optical depth ($\tau \lesssim 1$)
the driving mechanism becomes inefficient due to the weak coupling of matter and radiation.

In the outer layers of radiative envelopes, the timescale over which the fluid and radiation
exchange energy is generally much shorter than the typical mode periods $\sim 1/\omega_0$. In this
case the fluid and radiation have the same temperature $T$, and in the limit of rapid diffusion,
acoustic perturbations propagate at the isothermal sound speed
\begin{equation}
c_i^2 = \frac{p_{\rm gas}}{\rho},
\end{equation}
where $p_{\rm gas}$ is the gas pressure (in this regime, $c_i$ replaces $c_s$ 
in equation~\ref{eq:magnetosonic_disprel}). 
Quantitatively, the \emph{thermal locking} condition can be expressed as
\begin{equation}
\label{eq:thermal_locking_condition}
\omega_{\rm th} \equiv \dfrac{4(\gamma - 1)E}{p_{\rm gas}} \kappa_{\rm a} \rho c \gg \omega_0,
\end{equation}
where $\omega_{\rm th}$ is a heat exchange frequency (BS03), $E=a T^4$ is the radiation
energy density, $a$ is the radiation constant, 
$c$ is the speed of light, and $\kappa_{\rm a}$ is related to the
Planck-mean and Thomson scattering opacities (\S\ref{s:opacities}).

Assuming rapid diffusion (equation~\ref{eq:rapid_diffusion_condition}) and thermal 
locking (equation~\ref{eq:thermal_locking_condition}), the approximate RMI instability condition
for magnetosonic modes with wavelengths shorter than the gas pressure scale height
is (BS03, FS13)
\begin{eqnarray}
\label{eq:slow_instability_criterion}
\zeta\, F & \gtrsim & \zeta^4\left(p_{\rm gas} + \frac{4}{3}E \right)c_{\rm i} \phantom{abababab} [\textrm{slow modes}]\\
\label{eq:fast_instability_criterion}
\zeta^2 F & \gtrsim & \frac{1}{\zeta}\left(p_{\rm gas} + \frac{4}{3}E \right) v_{\rm A} \phantom{abababab} [\textrm{fast modes}]
\end{eqnarray}
where $F$ is the magnitude of the radiative flux $\mathbf{F}$,
\begin{equation}
\label{eq:zeta_def}
\zeta = \min\left( \frac{v_{\rm A}}{c_i},1\right),
\end{equation}
and the two equations refer to the slow- and fast magnetosonic 
branches in equation~(\ref{eq:magnetosonic_disprel}). Equations 
(\ref{eq:slow_instability_criterion})-(\ref{eq:fast_instability_criterion}) 
are obtained
from the exact dispersion relation of BS03 by ignoring angular factors involving
$\mathbf{k}$, $\mathbf{B}$, and $\mathbf{F}$. The two instability criteria
demand that driving by the background radiative flux overcome damping
by radiative diffusion. The factors of $\zeta$ are kept without simplification
to show the dependence of driving and damping terms on the magnetic field strength. The growth
rates are given by the difference between the left- and right
hand sides times a global prefactor independent of magnetic field ($\sim \kappa_{\rm F}/[c\,c_{\rm i}]$).

An additional driving mechanism exists in the short-wavelength regime when the
opacity has a density dependence. The instability criteria for
fast and slow modes are (BS03)
\begin{eqnarray}
\label{eq:opacity_instability_criterion_slow}
\zeta^3 F \Theta_\rho & \gtrsim & \zeta^4\left(\frac{4}{3}E + p_{\rm gas}\right) c_{\rm i}\phantom{abababab} [\textrm{slow modes}]\\
\label{eq:opacity_instability_criterion_fast}
F \Theta_\rho         & \gtrsim & \frac{1}{\zeta}\left(p_{\rm gas} + \frac{4}{3}E \right) v_{\rm A}\phantom{abababab} [\textrm{fast modes}], 
\end{eqnarray}
where
\begin{equation}
\label{eq:Theta_rho}
\Theta_\rho = \frac{\partial \ln \kappa_{\rm F}}{\partial \ln \rho}\bigg |_T.
\end{equation}
Equations~(\ref{eq:opacity_instability_criterion_slow})-(\ref{eq:opacity_instability_criterion_fast}) 
quantify the relative importance of driving due to the radiative flux acting on opacity variations versus
damping by radiative diffusion. As with the RMI instability criteria, the factors of $\zeta$ are 
not simplified so that driving and damping terms can be compared. The growth rates have the same
global prefactor as the RMI-driven case. RMI dominates driving of slow and fast modes when $\zeta^{-2}\gtrsim \Theta_\rho$
and $\zeta^2 \gtrsim \Theta_\rho$, respectively. In the limit of vanishing magnetic field ($v_{\rm A}\to 0$), the slow mode disappears,
and the fast mode becomes a purely acoustic mode that can only be destabilized at small scales 
by the radiative flux if $\Theta_\rho \neq 0$ (BS03).

\section{Methods}
\label{s:methods}

\subsection{Stellar Models}

We generate a grid of stellar models using the stellar evolution code MESA version 6794 
\citep{P2011,P2013,P2015,paxton_2017}. 
Since we are interested in stars with radiative envelopes with no density inversions, 
we choose an initial mass range $3-40\,M_\odot$
and adopt solar metallicity. Rotation is ignored for simplicity.
Models are considered up to the time at which they exhaust hydrogen at their centers, i.e., until
the end of the main sequence. The choice is motivated by the fact that stars spend most of their
time in this phase of evolution.

The quantities required to evaluate the instability criteria are 
taken directly from model profiles with the exception of the magnetic field, which is assumed and
not included in computing the stellar structure (\S\ref{s:magnetic_field}). 
We adopt three fiducial models with masses $6M_\odot$,
$12M_\odot$, and $30M_\odot$ for further analysis, corresponding to stars that end their life in
the Asymptotic Giant Branch, as a red-supergiant, and as a Wolf-Rayet star, respectively.

Most of the parameter choices for MESA models are the same as in \cite{fernandez_2017}: the \emph{Dutch} 
wind model \citep{dejager_1988,nugis_2000,vink_2001}, 
and the overshoot  choices of \citet{fuller_2015}. We use the tabulated opacities (Type 1) from the 
OPAL library \citep{rogers92,iglesias96}, and the `simple photosphere' option for the
atmospheric boundary condition \citep{P2011}.
The MESA {\tt inlist} files and {\tt run\_star\_extras.f} for the extraction of history columns are publicly 
available\footnote{{\tt https://bitbucket.org/rafernan/rmi\_mesa\_public}}. 

\subsection{Opacity}
\label{s:opacities}

The flux-mean opacity that enters the diffusion frequency (equation~\ref{eq:rapid_diffusion_condition}) 
is defined as (BS03)
\begin{equation}
\label{eq:flux_kappa_definition}
\kappa_{\rm F}\,F = \int_0^\infty d\nu\, \kappa^{\rm T}_\nu F_\nu
\end{equation}
where $F_\nu$ is the frequency-dependent flux, and 
\begin{equation}
\kappa^{\rm T}_\nu = \frac{1}{\rho},\left[\chi^{\rm th}(\rho,T_g) + n_e\sigma_{\rm T}\right]
\end{equation}
is the total transport opacity, with $\chi^{\rm th}$ the thermal absorption coefficient, 
$T_g$ the gas temperature, $n_e$ the electron density, and $\sigma_{\rm T}$ the Thomson scattering 
cross section.

In the diffusion approximation and in local thermodynamic
equilibrium (LTE), the flux-mean opacity is equal to the Rosseland mean opacity $\kappa_{\rm R}$ 
if the flux depends only on the spatial gradient of the radiation energy density (e.g., \citealt{huebner_2014}).
The assumption of LTE implies that each
layer of the star is in radiative equilibrium, therefore diffusion of photons in energy space should
be a secondary effect.

The absorption opacity $\kappa_a$ that enters the thermal locking frequency $\omega_{\rm th}$ is
defined as (BS03)
\begin{eqnarray}
\kappa_a & = & \kappa_{\rm P}\left( 1 + \frac{\partial\ln\kappa_{\rm P}}{\partial \ln T}\right) 
+ \frac{n_e \sigma_{\rm T}}{\rho}\frac{kT}{m_e c^2},\\
         & \simeq & \kappa_{\rm P}\left( 1 + \frac{\partial\ln\kappa_{\rm P}}{\partial \ln T}\right),
\end{eqnarray}
where $m_e$ is the electron mass, $k$ is Boltzmann's constant, and
we have assumed local thermodynamic equilibrium. The second equality
is valid when $kT \ll m_e c^2\simeq 0.5$~MeV, as is the case in the envelopes of massive stars.
The Planck-mean opacity is given by 
\begin{equation}
\label{eq:kappa_planck}
\kappa_{\rm P} = \frac{4\pi}{\rho\,a_\gamma cT^4}\int_0^\infty d\nu\,\chi^{\rm th}(\rho,T) \,B_\nu(T).
\end{equation}
This frequency average differs from $\kappa_{\rm R}$ in that (1) it excludes scattering
contributions, (2) the weighting favors frequencies at which absorption is the largest instead
of smallest, and (3) the weighting function peaks at a lower frequency for fixed temperature.
Whenever absorption dominates over scattering, $\kappa_{\rm P}$ is thus larger than 
$\kappa_{\rm R}$. This condition is safely satisfied in the radiative envelopes that are
the subject of our study. 

Given that MESA only outputs the Rosseland mean opacity and its derivatives, we compute
the thermal locking frequency with $\kappa_{\rm R}$ instead of $\kappa_{\rm P}$, which
yields a lower-limit on $\omega_{\rm th}$ when absorption dominates. To account for high-temperature regions
in which scattering dominates over absorption, we subtract the Thomson scattering
opacity from $\kappa_{\rm R}$, and set
\begin{equation}
\label{eq:kappa_a_effective}
\kappa_{\rm a} \simeq \kappa_{\rm R} - (1 + X)\frac{\sigma_{\rm T}}{2m_p},
\end{equation}
where $X$ is the hydrogen mass fraction\footnote{Equation~(\ref{eq:kappa_a_effective})
becomes negative in deep regions of the star for which scattering dominates, because
the OPAL opacities include corrections to the Thomson cross section due to special
relativity \citep{sampson59} and collective effects \citep{boercker87}. While we
could correct $\kappa_{\rm a}$ for these effects, the affected regions are deep enough
in the star that RMI effects are not important, thus we ignore this artifact.}.
Our results show that even with this conservative lower limit on $\omega_{\rm th}$,
the regions that are unstable to the RMI are well into the thermally-locked
regime. 

While the value $\omega_{\rm th}$ does not explicitly enter into the calculation
of the instability criteria, it determines the scale below which modes
thermally decouple and become adiabatic (BS03). The numerical results
of FS13 showed however that in the non-linear phase of the RMI, most of the power
is present at length scales comparable to the gas pressure scale height. 
We therefore ignore the dynamic range in wavelength over which the RMI
operates, and consider \emph{a posteriori} verification
of equation~(\ref{eq:thermal_locking_condition}) to be sufficient for the exploratory 
character of this work.

Finally, the density derivative of the flux-mean opacity $\Theta_\rho$ 
required to assess instability due to opacity variations 
(equations~\ref{eq:opacity_instability_criterion_slow}-\ref{eq:opacity_instability_criterion_fast})
is simply obtained from the density
derivative of the Rosseland-mean opacity.
\begin{equation}
\Theta_\rho = \frac{\partial \ln\kappa_{\rm R}}{\partial \ln\rho}\bigg |_T.
\end{equation}

\subsection{Magnetic Field and Instability Conditions}
\label{s:magnetic_field}

In order to evaluate the RMI instability criteria (equations~\ref{eq:slow_instability_criterion}
-\ref{eq:fast_instability_criterion}) we need to assume a magnitude and direction for the
magnetic field. For simplicity, we assume a constant background magnetic field inclined
$45^\circ$ relative to the outward radial direction. The fiducial magnitude is chosen
as $1$~kG following observed field strengths in O- to A-type stars \citep{grunhut_2017,sikora_2017}, 
and values of $0.3$ and $3$~kG are use to explore the sensitivity of results to this parameter.
A spatially uniform magnetic field is a good first approximation to a large-scale field that varies slowly
with depth in the outer envelope ($\sim $ outermost solar radius). 

It is also possible that sub-surface convection zones contribute to the
generation of localized magnetic fields via dynamo action (e.g.,
\citealt{cantiello_2011}). In this case, the magnetic field strength depends on
density as $B\propto \rho^{2/3}$ assuming that it rises buoyantly as a
spherical blob, and is normalized such that it is in equipartition with the
convective kinetic energy at the top of the iron convection zone.  We include
calculations that impose this spatial dependence of the field strength, in
order to test the sensitivity of our results to the field geometry.  \emph{We
note however that the instability conditions discussed in \S\ref{s:overview}
assume a uniform magnetic field (BS03), and hence they are formally valid when
$B/|dB/dr|\gg H_{\rm gas}$}, with $H_{\rm gas} = p_{\rm gas}/|dp_{\rm gas}/dr|$
the gas pressure scale height. Also, the mixing algorithms use to construct
our stellar models ignore rotation and magnetic fields.

The presence of an instability is inferred when any of the left hand sides of 
equations~(\ref{eq:slow_instability_criterion})-(\ref{eq:fast_instability_criterion})
are larger than their respective right hand sides. Based on the results of FS03, we set
$k = 1/H_{\rm gas}$ in all calculations. The operation of any of these instabilities 
is contingent on the rapid diffusion condition (\ref{eq:rapid_diffusion_condition}) being satisfied, 
thus it is also evaluated. Finally, the condition of thermal locking (\ref{eq:thermal_locking_condition}) 
must be well satisfied for the validity of the equations used.

\section{Results}
\label{s:results}

\begin{figure*}
\includegraphics*[width=\textwidth]{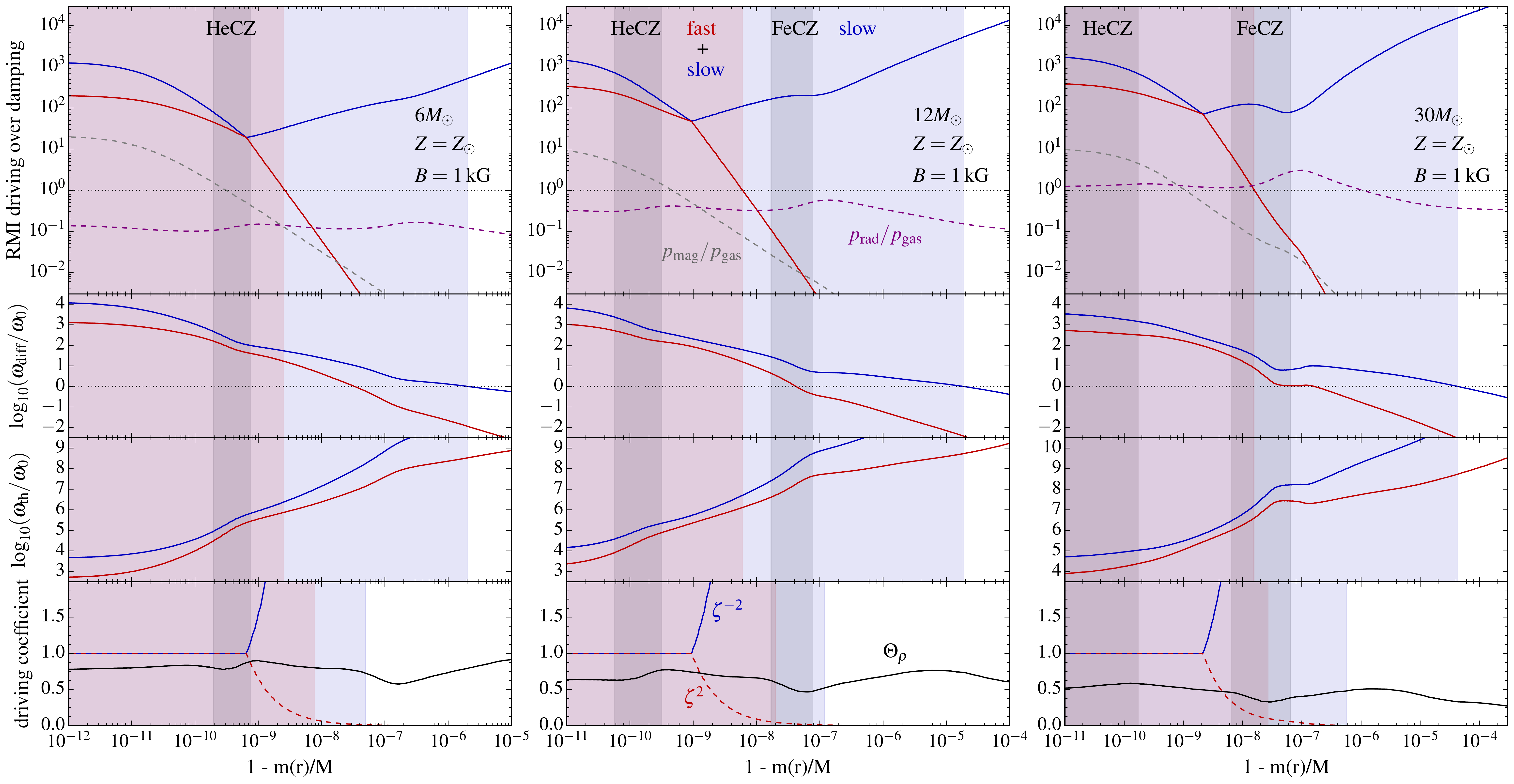}
\caption{
RMI instability regions as a function of fractional mass from the stellar surface, for 3 fiducial solar-metallicity
models with masses $6M_\odot$ (left), $12M_\odot$ (center), and $30M_\odot$ (right), assuming a uniform magnetic
field of $1$~kG. The time is chosen to be $1/2$ of the main sequence lifetime for each model (c.f. Figure~\ref{f:evolution}).
\emph{Top row:} Ratio of RMI driving to damping terms (left over right sides of 
equations~\ref{eq:slow_instability_criterion}-\ref{eq:fast_instability_criterion})
for slow (solid blue) and fast (solid red) magnetosonic modes. The blue shaded area shows the region
of the star in which only slow modes are unstable, whereas the red shaded area denotes zones where both slow and fast 
modes are unstable. The gray
shaded areas correspond to convection zones driven by helium (left) and iron (right) opacity peaks.
The purple and gray dashed lines show the ratios of radiation and magnetic to gas pressures, respectively.
\emph{Second row:} ratio of diffusion frequency to mode frequency for slow (blue) and fast (red) magnetosonic
modes (equation~\ref{eq:rapid_diffusion_condition}). \emph{Third row:} ratio of heat exchange frequency
between gas and radiation to the mode frequency, for slow (blue) and fast (red) magnetosonic modes 
(equation~\ref{eq:thermal_locking_condition}).
\emph{Bottom row:} Coefficients that determine the relative importance of driving of short-wavelength modes 
by the RMI and by the radiative flux acting on opacity
variations. The RMI dominates driving of slow and fast modes when $\zeta^{-2}\gtrsim \Theta_\rho$ and 
$\zeta^2 \gtrsim \Theta_\rho$, respectively, where $\zeta$ is given by equation (\ref{eq:zeta_def}) and 
$\Theta_\rho$ by equation~(\ref{eq:Theta_rho}). The shaded areas in this panel show the regions in 
which slow and fast modes are unstable to driving by opacity variations 
(equations~\ref{eq:opacity_instability_criterion_slow}-\ref{eq:opacity_instability_criterion_fast}).}
\label{f:profiles}
\end{figure*}

\begin{figure*}
\includegraphics*[width=\textwidth]{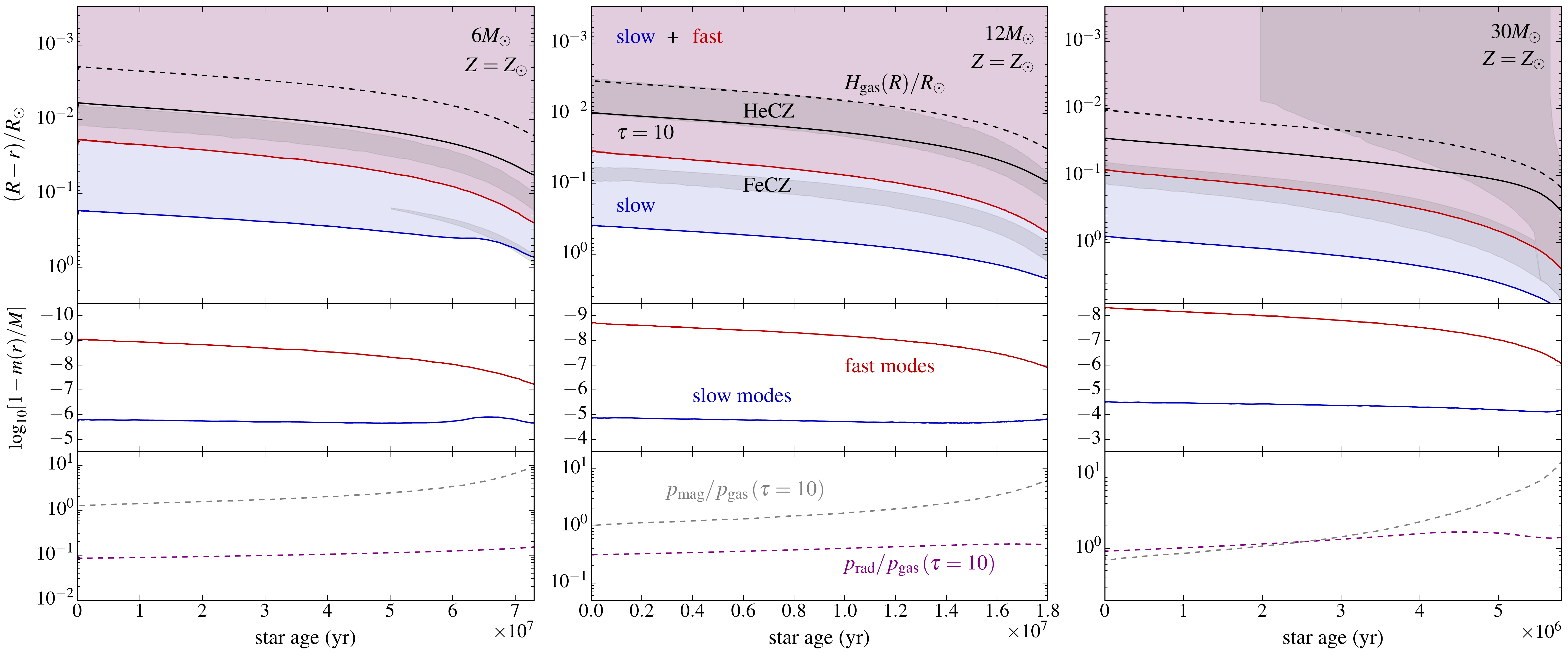}
\caption{
Instability regions as a function of time in the main sequence, for the three fiducial
solar metallicity models ($6M_\odot$, $12M_\odot$, and $30M_\odot$) and a 
uniform magnetic field of $1$~kG. \emph{Top row:}
depth below the stellar surface up to which only slow modes (blue curve) 
and fast plus slow modes are unstable (red curve). The gray shaded bands
correspond to convection zones at the helium (top) and iron (bottom) opacity 
peaks. The black solid line denotes the radius at which the optical depth from the surface is 
$\tau=10$, and the black dashed line shows the radius one gas pressure scale height from the surface.
\emph{Middle row:} Fractional mass from the stellar surface where slow 
modes (blue) and fast modes (red) become unstable. \emph{Bottom row:} ratio of
radiation to gas pressure (purple dashed) and magnetic to gas pressure (gray
dashed) at the radius at which $\tau = 10$.
}
\label{f:evolution}
\end{figure*}

\subsection{Instability regions for selected stars}
\label{s:profiles}

The spatial locations of the RMI-unstable regions in the three fiducial
models ($6$, $12$, and $30M_\odot$ at solar metallicity) with spatially uniform
magnetic field are shown in Figure~\ref{f:profiles}. All three stars display 
the same qualitative features: (1) the outermost part of the envelope 
is unstable to both fast and slow magnetosonic waves, (2) the slow modes are unstable
into deeper stellar layers than the fast modes, and (3) the instability
regions overlap with the helium- and iron opacity peak convection zones.

The instability region for slow modes is set by the
rapid diffusion condition (equation~\ref{eq:rapid_diffusion_condition}).
This is partly a consequence of our assumption of constant magnetic field,
which causes the ratio of magnetic to gas pressure $p_{\rm mag}/p_{\rm gas}$
to drop with increasing depth in the star [$p_{\rm mag} = B^2/(8\pi) = \rho v_{\rm A}^2/2$ is the magnetic
pressure]. Damping due to radiative diffusion for
this mode is suppressed relative to driving by a factor $(v_{\rm A}/c_{\rm i})^3$ 
when $v_{\rm A}< c_{\rm i}$, thus radiative driving of this mode dominates throughout the star. 
When the rapid diffusion condition is no longer satisfied, however, magnetosonic modes
become adiabatic and no driving occurs.

Fast magnetosonic waves become unstable closer to the stellar surface
than slow modes because damping by radiative diffusion is more efficient, and therefore
a higher radiative forcing is required than for slow modes at the same depth. FS13 found that fast modes become
unstable when a parameter measuring the diffusion speed $c/\tau_0$ relative to the
isothermal sound speed $c_i$ exceeds unity
\begin{equation}
\Re = \frac{\ell_{\rm H}}{4}\frac{c/c_{\rm i}}{\tau_0} > 1
\end{equation}
where $\tau_0 = \kappa_{\rm F}\rho H_{\rm g}$ is the optical depth over
a gas pressure scale height, and 
\begin{equation}
\ell_{\rm H} = \frac{H_{\rm gas}}{H_{\rm rad}} = \frac{p_{\rm gas}}{p_{\rm rad}}\bigg|\frac{dp_{\rm rad}/dr}{dp_{\rm gas}/dr}\bigg|
\end{equation}
is the ratio of gas to radiation pressure scale heights (which is of order unity),
with $p_{\rm rad} = E/3$ the radiation pressure.
The parameter $\Re$ increases close to the stellar surface given that the isothermal sound speed
and the density decrease sharply, and because the ratio of flux to radiation energy density
\begin{equation}
\frac{F}{Ec}\sim \frac{1}{\tau_0}
\end{equation}
also increases. Because driving depends on the radiative flux and damping on 
radiative diffusion and thus on $E$ (equations~\ref{eq:slow_instability_criterion}-\ref{eq:fast_instability_criterion}),
the ratio of $F$ to $E c$ and thereby $\Re$ quantify
the amount of free-energy available to drive the RMI (BS03).
Fast modes are therefore guaranteed to become unstable close enough to the photosphere. 
The fact that the RMI is valid in the diffusion approximation means, however, that there is a 
lower limit in optical depth to this instability region.

Figure~\ref{f:profiles} shows that in all models, both magnetosonic branches are
well within the thermally coupled regime, with $\omega_{\rm th}\gtrsim 10^3\omega_0$ at 
the stellar surface. We therefore confirm \emph{a posteriori} the validity of the
instability equations used. The use of the Rosseland mean opacity in computing
$\omega_{\rm th}$ means that the real thermal coupling is likely to be much 
larger in regions where electron scattering does not dominate the opacity (\S\ref{s:opacities}).

Higher mass stars have a higher ratio of radiation to gas pressure throughout
their interior. The larger importance of radiation pressure is associated with 
stronger radiative forcing of fast and slow modes for higher mass stars at any given
point in the envelope, resulting in deeper instability regions for higher mass stars.
Going from $6M_\odot$ to $30M_\odot$, the exterior mass unstable to slow modes
increases from $\sim 10^{-6}$ to almost $10^{-4}$ of the total stellar mass, while the 
fast modes affect the outermost $10^{-9}-10^{-8}$ of the mass.

Figure~\ref{f:evolution} shows the evolution on the main sequence of the three fiducial models with 
constant magnetic field up to the time at which central hydrogen is exhausted. 
Over this period of time, stars expand in radius by a factor of a few, increasing in luminosity
and decreasing in surface temperature. The instability regions move deeper in both
radius and mass as the star evolves, approximately tracking surfaces of constant
temperature, as shown by their relation to the convection zones at opacity peaks,
which occur at specific temperatures.

The radial extent of the instability regions is also shown in Figure~\ref{f:evolution}.
Given that the RMI is valid in the optically thick regime, we adopt a fiducial value 
of the optical depth from the surface $\tau=10$ as a conservative boundary of the
region inside which the diffusion approximation is valid. Figure~\ref{f:evolution}
shows that this radius lies a few gas pressure scale heights inside the stellar
surface. 

If the RMI is to serve as a source of sub-photospheric fluctuations,
the place in the star where it is expected to be most prominent is where the
free energy for the instability is the largest  ($\Re\gg 1$) while being simultaneously in the 
optically-thick regime, i.e., where the optical depth is not much larger than $1$. The ratios of 
radiation to gas pressures and magnetic to gas pressures at $\tau=10$ are shown in Figure~\ref{f:evolution}.
For a uniform field of $1$\,kG, magnetic and gas pressures are of the same
order except near the end of the main sequence. The ratio of radiation to
gas pressure does not depend strongly on depth and instead varies strongly with stellar mass.

In all 3 fiducial models with a uniform magnetic field, both fast and slow modes are unstable 
inside the radius where $\tau=10$, and
therefore can be considered to operate in the parameter regime where the RMI is valid
and most effective. For fast modes, the spatial region involved ranges from a few percent to
$\sim 10\%$ of a solar radius for the $6M_\odot$ model, to $0.1-2R_\odot$ for the $30M_\odot$ star.
The region where slow modes are unstable is much larger, ranging from $\sim 0.1R_\odot$ at a minimum for the $6M_\odot$
to several solar radii at the end of the main sequence lifetime of the $30M_\odot$ star.

\begin{figure*}
\includegraphics*[width=\textwidth]{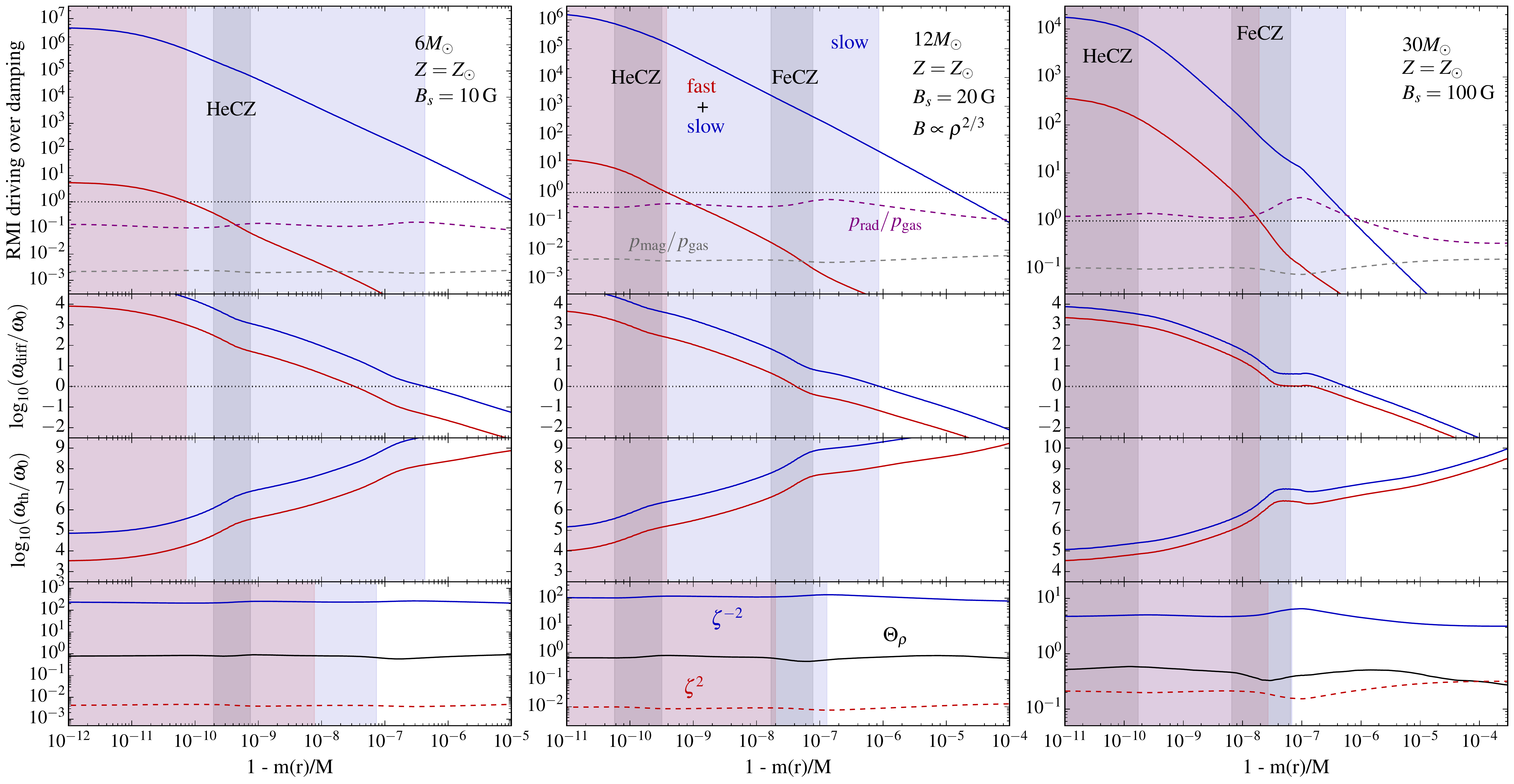}
\caption{
Same as Figure~\ref{f:profiles}, but now assuming a magnetic field strength that scales with density
as $B\propto \rho^{2/3}$, as appropriate for flux freezing in a buoyantly rising spherical blob. 
For the $12M_\odot$ and $30M_\odot$ models, the field is normalized so that the magnetic energy density 
is equal to the peak kinetic energy density in the iron convection zone (c.f. \citealt{cantiello_2011}).
The resulting surface fields $B_s$ are labeled in each panel. For the $6M_\odot$ star, the iron
convection zone is non-existent or too weak when it appears later in the evolution, thus we simply impose a surface
magnetic field of $10$\,G for comparison.
}
\label{f:profiles_flux-freeze}
\end{figure*}

\subsubsection{Instability regions for a radially-varying magnetic field}

In order to explore how the instability regions depend on the amplitude and radial
dependence of the magnetic field strength, we explore a scenario in which the
field is generated by dynamo action in the iron convection zone. 
 In this case, flux-freezing on buoyantly-rising
blobs implies a field strength scaling $B\propto \rho^{2/3}$, and a normalization such
that the magnetic energy density matches the kinetic energy density in the convection
zone \citep{cantiello_2011}.

Figure~\ref{f:profiles_flux-freeze} shows the instability regions with this field
configuration, for the same fiducial models shown in Figure~\ref{f:profiles}. The
resulting fields at the surface are $20$\,G and $100$\,G for the $12M_\odot$ and
$30M_\odot$ models, respectively. In the case of the $6M_\odot$ star, the convection zone is too
weak to generate any interesting field so we arbitrarily impose a surface field of $10$\,G to
compare with the other models.

The instability regions shrink in mass, approximately by a factor $10$. Nevertheless,
slow modes are still driven by the RMI to a depth comparable to that of the iron convection
zone. Regarding driving strength, the slow mode growth rate is suppressed by a factor $(v_{\rm A}/c_{\rm i})$
when $v_{\rm A} < c_{\rm i}$. The smaller ratio of magnetic to gas pressure relative
to the case of a slowly-varying $\sim $\,kG field means that
the efficiency with which energy from the radiation field is tapped by the RMI
is low in this case.

While fast modes are technically unstable to the RMI to depths larger than $\tau=10$
for the $12M_\odot$ and $30M_\odot$ stars, the driving is negligible compared
to that due to the flux acting on opacity variations (\ref{s:opacity_variations}).

We emphasize again that the instability conditions derived by BS03 assume
a spatially uniform magnetic field, and are therefore valid only on
spatial scales comparable or smaller than the gas pressure scale height.

\subsubsection{Relation between driving due to the RMI and density-dependent opacity}
\label{s:opacity_variations}

In the limit of small damping by radiative diffusion, the relative importance of
the RMI and driving from opacity variations is given
by the ratio of the terms on the left hand sides of
equations~(\ref{eq:slow_instability_criterion})-(\ref{eq:fast_instability_criterion}) 
and (\ref{eq:opacity_instability_criterion_slow})-(\ref{eq:opacity_instability_criterion_fast}), respectively
for each mode. The RMI dominates the driving of slow modes
if $\zeta^{-2} > \Theta_\rho$ and of fast modes if $\zeta^2 > \Theta_\rho$. Note that
these conditions are independent of the ratio of radiation to gas pressure, since the
flux scales out.

The bottom panel of Figure~\ref{f:profiles} shows these factors for the three fiducial models
with uniform $1$\,kG magnetic field. In addition, the regions that are unstable to driving
due to opacity variations 
(equations~\ref{eq:opacity_instability_criterion_slow}-\ref{eq:opacity_instability_criterion_fast}) 
are shown as shaded regions following the same color-coding as
the RMI-driven regions in the panels above.
Since at this field strength $p_{\rm mag}\gtrsim p_{\rm gas}$
at $\tau = 10$ for all three models, the ratio of driving terms is $\Theta_\rho^{-1}>1$ and the RMI dominates
over opacity fluctuations by a factor of a few. Moving inward in depth, for a constant magnetic 
field, the importance of magnetic pressure relative to gas pressure decreases, and therefore $\zeta$
also decreases with depth. 

\begin{figure}
\includegraphics*[width=\columnwidth]{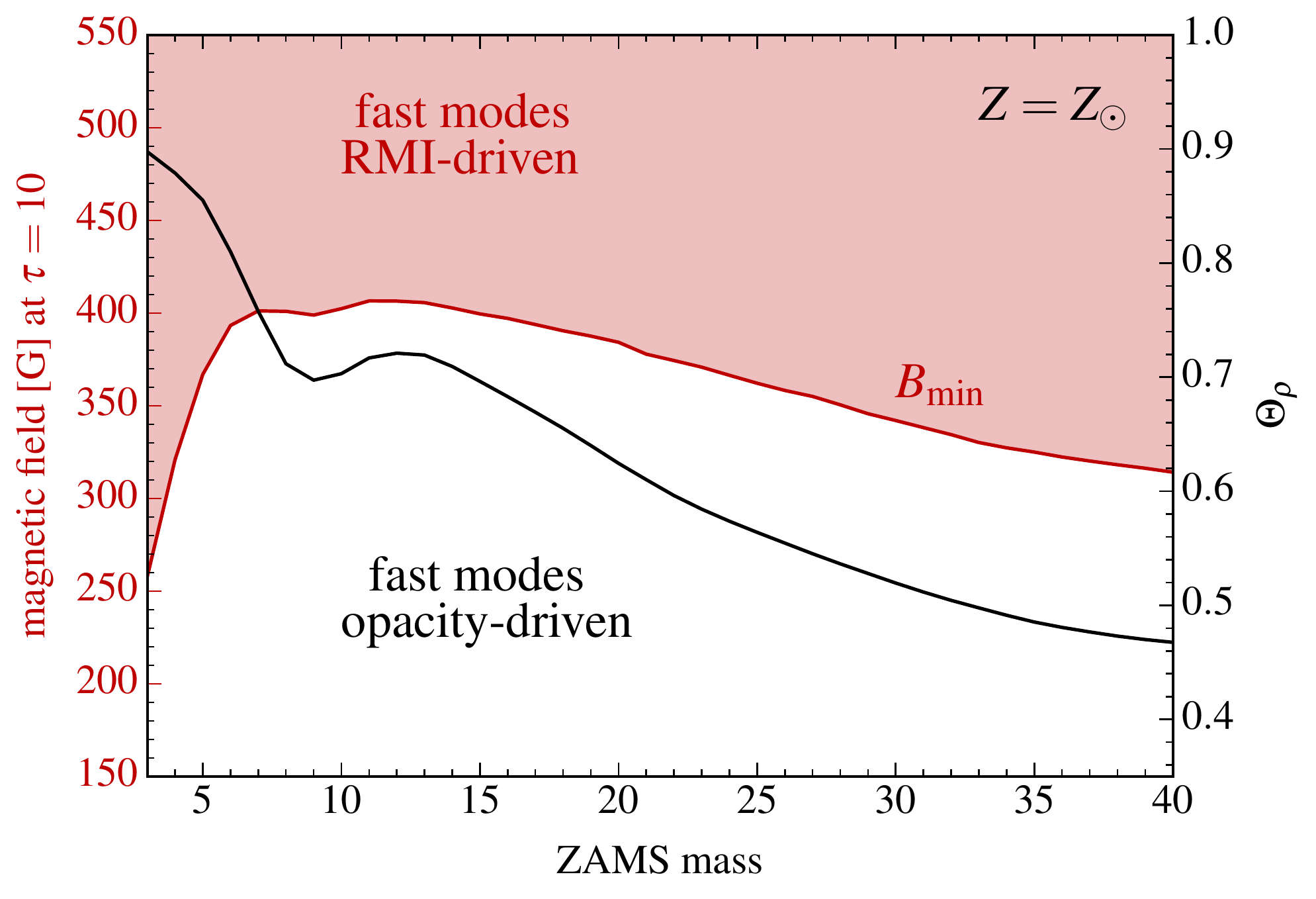}
\caption{
\emph{Left:} minimum magnetic field for dominance of the RMI in driving of 
short-wavelength fast magnetosonic modes (equation~\ref{eq:bmin}, red curve) 
evaluated at $\tau =10$, as a function of ZAMS mass for our grid of solar metallicity models. The time
corresponds to that at which 
the central hydrogen abundance has decreased to one half of its initial value.
\emph{Right:} logarithmic opacity derivative (equation~\ref{eq:Theta_rho}) at $\tau = 10$ for the same models at the same times.
}
\label{f:bmin}
\end{figure}

Given the conditions in the envelopes of these fiducial models, RMI driving of slow modes always 
dominates over that due to opacity-variations. This condition is independent of the magnetic field strength,
relying instead on the condition $\Theta_\rho < 1$, which is satisfied in all of our models (Figure~\ref{f:bmin})
but can depend on other variables such as metallicity. The dominance of slow modes arises
because driving by the flux acting on opacity variations has the same suppression factor $(v_{\rm A}/c_{\rm i})^3$ 
as damping by radiative diffusion when $v_{\rm A} < c_{\rm i}$. This suppression factor 
is associated with the compressional component of the wave. Despite this dominance, 
the growth rate of slow modes is suppressed by a factor $v_{\rm A}/c_{\rm i}$ in the 
limit of weak damping, therefore for a constant magnetic field, RMI activity should decrease 
with depth.

In the case of fast modes, RMI-driving becomes sub-dominant once 
$v_{\rm A}/c_{\rm i}\lesssim \Theta_\rho^{1/2}$. In this case, upward-propagating 
($\mathbf{\hat k}\cdot\mathbf{F}>0$) fast modes are driven by the flux acting
on opacity variations, and downward going fast modes are damped (BS03). The instability
region for fast modes driven by opacity variations is deeper than that due to the
RMI, extending all the way to the top of the iron convection zone for the $12M_\odot$
and $30M_\odot$ models.

By setting a minimum optical depth at which the RMI can operate (given by the applicability of 
the diffusion approximation), the condition $v_{\rm A}/c_{\rm i}\geq \Theta_\rho^{1/2}$ sets 
a minimum magnetic field for which driving of fast modes is dominated by the RMI:
\begin{equation}
\label{eq:bmin}
B_{\rm min} = \left({4\pi p_{\rm gas} \Theta_\rho}\right)^{1/2} = \left(\frac{\Theta_\rho}{2}\right)^{1/2}\,B_{\rm eq}
\end{equation}
where $B_{\rm eq} = \sqrt{8\pi p_{\rm gas}}$ is the magnetic field that results 
in equipartition of magnetic pressure with gas pressure.

Figure~\ref{f:bmin} shows $B_{\rm min}$ at $\tau = 10$ as a function of initial 
stellar mass for all our models. For a ZAMS $M<6M_\odot$, this minimum field increases due to an increase in the stellar pressure 
near the surface. The field plateaus at about $400$\,G for stars in the mass range $6-15M_\odot$,
subsequently decreasing at higher masses mostly due to a drop in $\Theta_\rho$ with increasing
stellar mass. For fields significantly lower than $\sim 300$\,G, fast magnetosonic modes
are driven unstable primarily by the radiative flux acting on opacity variations, for all stellar masses.

The relative hierarchy of driving mechanisms in the weak magnetic field regime is illustrated
in the bottom panel of Figure~\ref{f:profiles_flux-freeze}. Here, the condition $p_{\rm mag}\ll p_{\rm gas}$
results in $\zeta^2\ll \Theta_\rho \ll \zeta^{-2}$, with slow modes driven by the RMI and
fast modes driven by the flux acting on opacity variations. As in the strongly magnetized
case, the region in which fast modes are driven by opacity variations extends all the way
to the surface of the iron convection zone for the $12M_\odot$ and $30M_\odot$ models.

\subsubsection{Metallicity effects}
\label{s:metallicity}

\begin{figure}
\includegraphics*[width=\columnwidth]{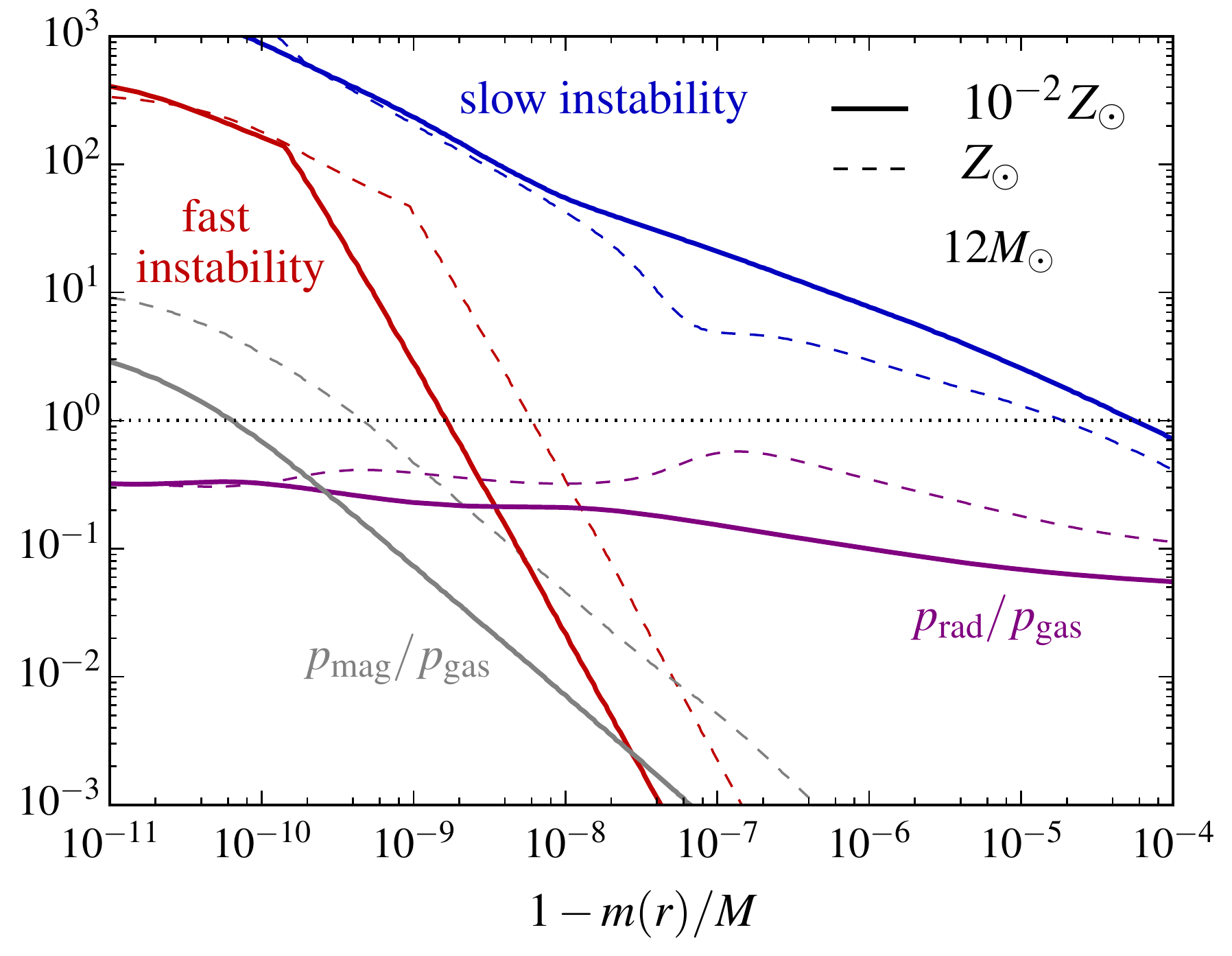}
\caption{Profiles of RMI-related quantities for two $12M_\odot$ models, one with 
solar metallicity (dashed lines) and another with metallicity $10^{-2}$ times solar (solid lines).
The time shown corresponds approximately to one half of the main sequence lifetime for
each model.
The `slow instability' curve shows $\omega_{\rm diff}/\omega_{\rm 0}$ for 
slow modes (equation~\ref{eq:rapid_diffusion_condition}), and `fast instability' corresponds to 
the ratio of left to right hand sides of equation~(\ref{eq:fast_instability_criterion}).
Compare with Figure~\ref{f:profiles}.}
\label{f:metallicity}
\end{figure}

To investigate the effects of metallicity on the RMI, we evolve a $12M_\odot$ model
with $Z=10^{-2}Z_\odot$. During the main sequence, the model loses less mass than
the solar metallicity model, has a smaller radius (by $\sim 50\%$) and higher 
effective temperature (by $\sim 30\%$) at all times. The iron convection zone is absent for most
of the main sequence evolution, and the helium convection zone has a smaller convective
Mach number than the solar metallicity model at a given age.

The unstable magnetosonic regions are qualitatively the same as
in the solar metallicity model, as shown in Figure~\ref{f:metallicity}. For a uniform $1$\,kG magnetic 
field and at a time equal to one half of the main sequence lifetime, the ratio of magnetic to 
gas pressure is smaller by a factor $\sim 3$ throughout the star compared to the solar
metallicity model, while the ratio of radiation to gas pressure can be smaller by a factor up to $\sim 2$.
The region where slow modes are unstable extends to higher depths than in the solar 
metallicity model. This is expected for a lower opacity, which increases the diffusion 
frequency (equation~\ref{eq:rapid_diffusion_condition}). The region where fast modes
are unstable extends to a shallower depth. This is consistent with stronger damping by radiative
diffusion whenever $v_{\rm A} < c_i$ (equation~\ref{eq:fast_instability_criterion}).

The values of the ratios $p_{\rm rad}/p_{\rm gas}$ 
and $p_{\rm mag}/p_{\rm gas}$ are comparable in both models, and therefore 
the instability operates in the same regime in both cases. We therefore do not expect the 
instability to be very sensitive to metallicity for the same background magnetic field.

\subsection{Relation to convection zones from opacity peaks}

\begin{figure}
\includegraphics*[width=\columnwidth]{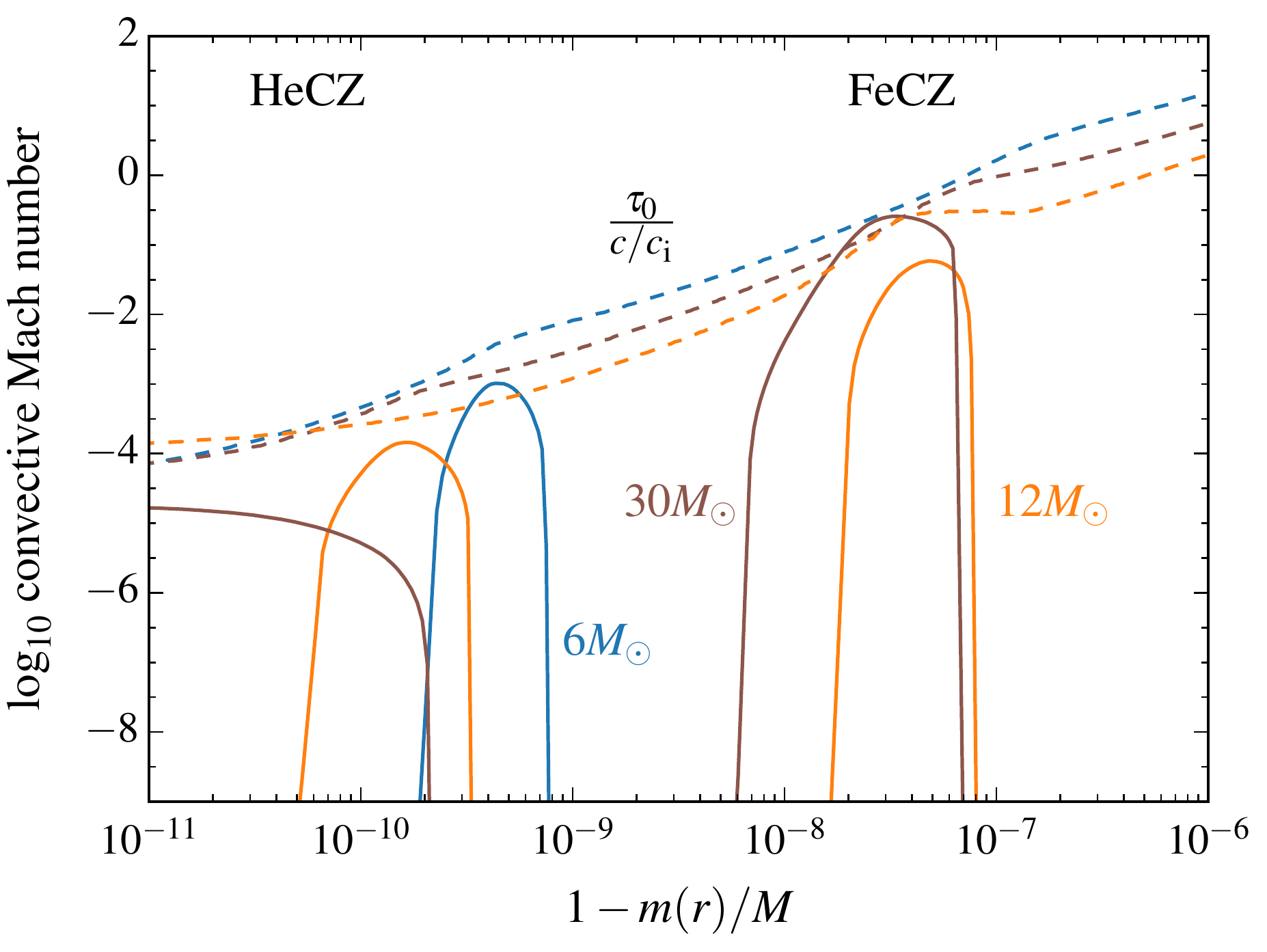}
\caption{
Convective Mach number at the time where the central hydrogen mass fraction is $0.35$ for
the three fiducial models (solid lines). Also shown as dashed lines is the convective efficiency parameter 
$\tau_0/(c/c_{\rm i})$, where $\tau_0$ is the flux-mean optical depth over a gas pressure scale height.
}
\label{f:conv_efficiency}
\end{figure}

Convection zones due to helium- and iron opacity peaks 
are ubiquitous in our stellar models, as shown
in Figures~\ref{f:profiles}-\ref{f:profiles_flux-freeze}. Both convection
zones are completely contained by the region in which slow magnetosonic
modes are unstable to the RMI, and the iron convection zone can sometimes overlap
the region where fast modes are unstable to the RMI, as in the $30M_\odot$ model.
It is therefore useful to clarify how these two very different sources of density
fluctuations interplay with each other.

In a convectively unstable region, the relative importance of radiation diffusion 
and convection in transporting energy is determined by the convective efficiency.
This efficiency can be quantified by the ratio of the
isothermal sound speed $c_i$ to the diffusion speed $c/\tau_0$. This ratio 
was found by \citet{jiang_2015} to correctly describe the regimes of high and low convective
efficiency ($c_{\rm i} \gg c/\tau_0$ and $c_{\rm i} \ll c/\tau_0$, respectively)
in radiation-hydrodynamic simulations of opacity-driven convection zones.

Figure~\ref{f:conv_efficiency} shows the convective Mach number in the opacity-driven
convection zones halfway through the main-sequence time for our fiducial models (same times
shown in Figure~\ref{f:profiles}). As previously described by \citet{C09}, 
convective energy transport at the helium opacity peak is very inefficient, and 
can therefore be ignored. At the iron opacity peak, the convective Mach numbers
are larger, and the efficiency is at most of the order of $10\%$ at the end of the
main sequence for the $30M_\odot$ star. The fraction of the stellar luminosity
carried by this convection zone is at most a few percent in our models,
although the Mach numbers can result in significant gravity wave excitation (the
subsonic character of these convection zones make them less efficient
at exciting acoustic waves directly; \citealt{C09}).

\begin{figure}
\includegraphics*[width=\columnwidth]{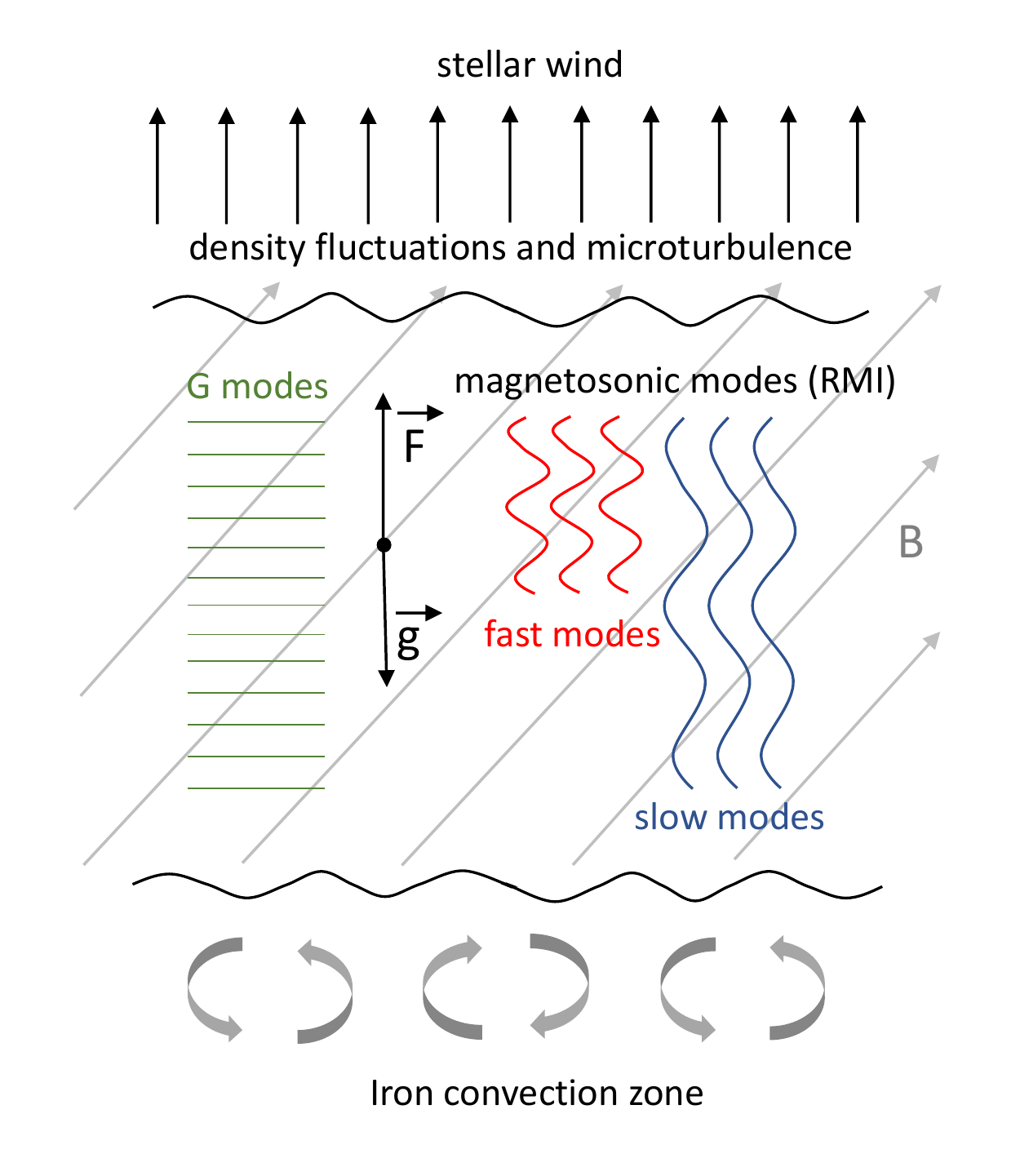}
\caption{Schematic diagram of the operation of the RMI in massive star envelopes,
with the vertical direction along the radial coordinate ($\mathbf{g}$ is the
acceleration of gravity and $\mathbf{F}$ is the background radiative flux).
The star is radiative between the upper edge of the near-surface convection zone
-- driven by the iron opacity peak -- and the photosphere. 
Compressional, short-wavelength MHD waves (slow and fast magnetosonic modes) are destabilized by the radiation flux 
in the radiative zone when a magnetic field $\mathbf{B}$ is present. 
Since convection is subsonic, it is most effective at exciting gravity waves into 
the radiative layer. Perturbations from RMI-destabilized magnetosonic modes and convectively-excited
gravity waves propagate toward the photosphere, providing a source of fluctuations
that can result in microturbulence and wind clumping.
}
\label{f:schematics}
\end{figure}

We therefore consider an overall picture in which the iron convection
zone excites gravity waves, as proposed by \citet{C09}, and the RMI excites compressional MHD
waves in the radiative zone in between this convection zone and 
the photosphere (Figure~\ref{f:schematics}). For stars at the low
end of the mass range for radiative envelopes, the iron convection zone appears late in 
the main sequence (e.g., the $6M_\odot$ model in Figure~\ref{f:evolution}), 
thus the RMI (or opacity variations at low magnetic field) may be the only source of 
small-scale sub-photospheric fluctuations. This could also be the case for stars at very low
metallicity, for which iron convection zones are also weaker. For larger masses, 
and/or higher metallicities, convection zones become increasingly stronger 
and contribute with a larger energy flux in gravity waves. The interplay between
the two mechanisms as a function of stellar mass is explored quantitatively
in the following subsection.

\subsection{Dependence on stellar mass}
\label{s:mass}

\begin{figure}
\includegraphics*[width=\columnwidth]{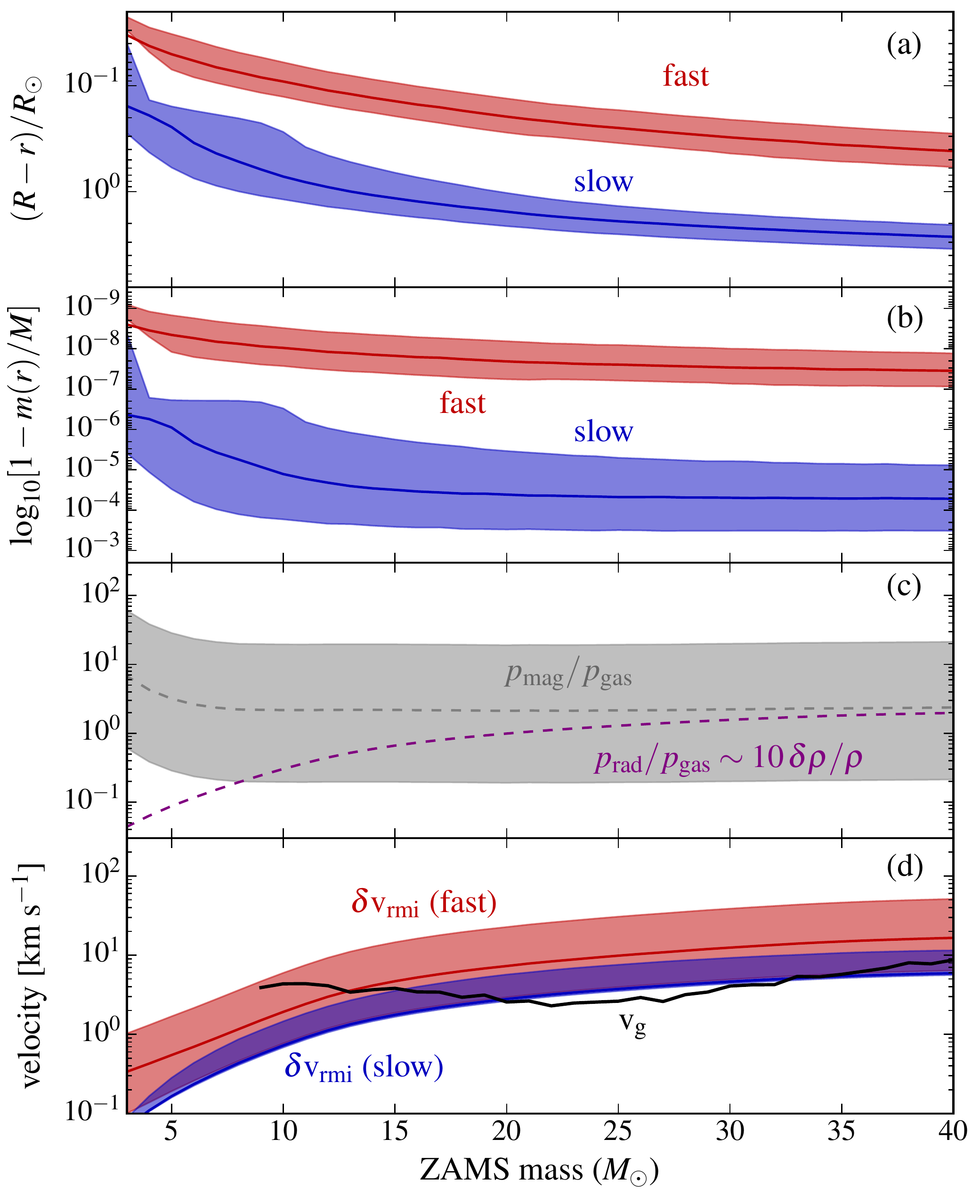}
\caption{
Properties of the RMI as a function of initial stellar mass, for
solar metallicity models with a uniform 1\,kG magnetic field. Shaded areas show the 
variation obtained for field strengths in the range $300$\,G to $3$\,kG. 
\emph{Top:} Radial depth (in $R_\odot$) from the
surface where fast (red) and slow (blue) magnetosonic modes are unstable.
\emph{Second panel:} Exterior mass unstable to the RMI, with fast and slow modes 
shown by the red and blue curve, respectively. \emph{Third panel:} Ratio of radiation 
to gas pressure (purple) and magnetic to gas pressure (gray) at a radius where the
optical depth from the surface is $\tau = 10$. For reference, the fractional density
perturbations obtained in the simulations of FS13 satisfy approximately 
$\delta\rho/\rho \sim 0.1\,p_{\rm rad}/p_{\rm gas}$ (equation~\ref{eq:drho_rmi}). 
\emph{Bottom:} Estimated velocity perturbation at $\tau = 10$ due to fast (red) 
and slow (blue) magnetosonic modes obtained from equation~(\ref{eq:dv_rmi}).
For reference, we also show the estimated velocity perturbation at $\tau = 10$
from gravity waves excited by the iron convection zone (equation~\ref{eq:vg_cantiello}).
}
\label{f:zams}
\end{figure}

Figure~\ref{f:zams} shows the location of the RMI instability regions for slow- and fast modes
as a function of initial stellar mass for all our models ($3-40M_\odot$) with a constant
magnetic field of $1$\,kG. The age of the star is that at which the central hydrogen mass 
fraction decreases to $50\%$ of its initial value.

The relative hierarchy of slow- and fast mode unstable regions is maintained for
all models. The instability regions extend into deeper fractional mass and radii
for more massive stars. 
The ratio of radiation to gas pressure at $\tau = 10$ is a monotonically increasing
function of ZAMS mass, ranging from $p_{\rm rad}/p_{\rm gas} \simeq 0.04$ at $3M_\odot$
to $p_{\rm rad}/p_{\rm gas}\simeq 2$ at $40M_\odot$. The radiation pressure is
at most comparable to the gas pressure. For the imposed
magnetic field of 1\,kG, the ratio of magnetic to gas pressure ranges from
$p_{\rm mag}/p_{\rm gas}\simeq 7$ at $3M_\odot$ to $p_{\rm mag}/p_{\rm gas}\simeq 2$
at $40M_\odot$. In other words, the relevant sub-photospheric region is 
in the regime $p_{\rm mag}\sim p_{\rm gas}$, for which the
RMI saturation amplitudes are maximized (FS13). Note however that as time elapses in the
main sequence, the magnetic-to-gas pressure at $\tau = 10$ increases if the
magnetic field is assumed to be constant (Figure~\ref{f:evolution}).

The simulations of the nonlinear development of the RMI by FS13 found saturation amplitudes 
that roughly satisfy:
\begin{equation}
\label{eq:drho_rmi}
\frac{\delta \rho}{\rho}\sim 0.1\frac{p_{\rm rad}}{p_{\rm gas}}
\end{equation}
for $p_{\rm rad}\lesssim p_{\rm gas}$, where $\delta \rho/\rho$ is the 
root-mean-square density fluctuation relative to the uniform initial background.
Based on this expression we can relate the velocity perturbation associated
with the RMI to the ratio of radiation to gas pressure. The
velocity perturbation in the linear phase is (e.g. Appendix A of FS13)
\begin{equation}
\label{eq:dv_rmi}
\delta v_{\rm rmi}\simeq v_{\rm ph}\,\frac{\delta \rho}{\rho}
  \left[\frac{2v_{\rm ph}^2 - c_{\rm i}^2 - v_{\rm A}^2}{v_{\rm ph}^2 - (\mathbf{\hat k}\cdot \mathbf{v}_{\rm A})^2} \right]^{1/2}
\end{equation}
The term in square brackets is of order unity except for slow modes in the limit $v_{\rm A}\ll c_{\rm i}$, in 
which case it becomes $(c_{\rm i}/v_{\rm A})^2$, or $\zeta^{-2}$. Given the moderate
amplitudes implied by equation~(\ref{eq:drho_rmi}), this expression is a reasonable
first estimate for the velocity fluctuations induced by the RMI.

\begin{figure*}
\includegraphics*[width=0.945\columnwidth]{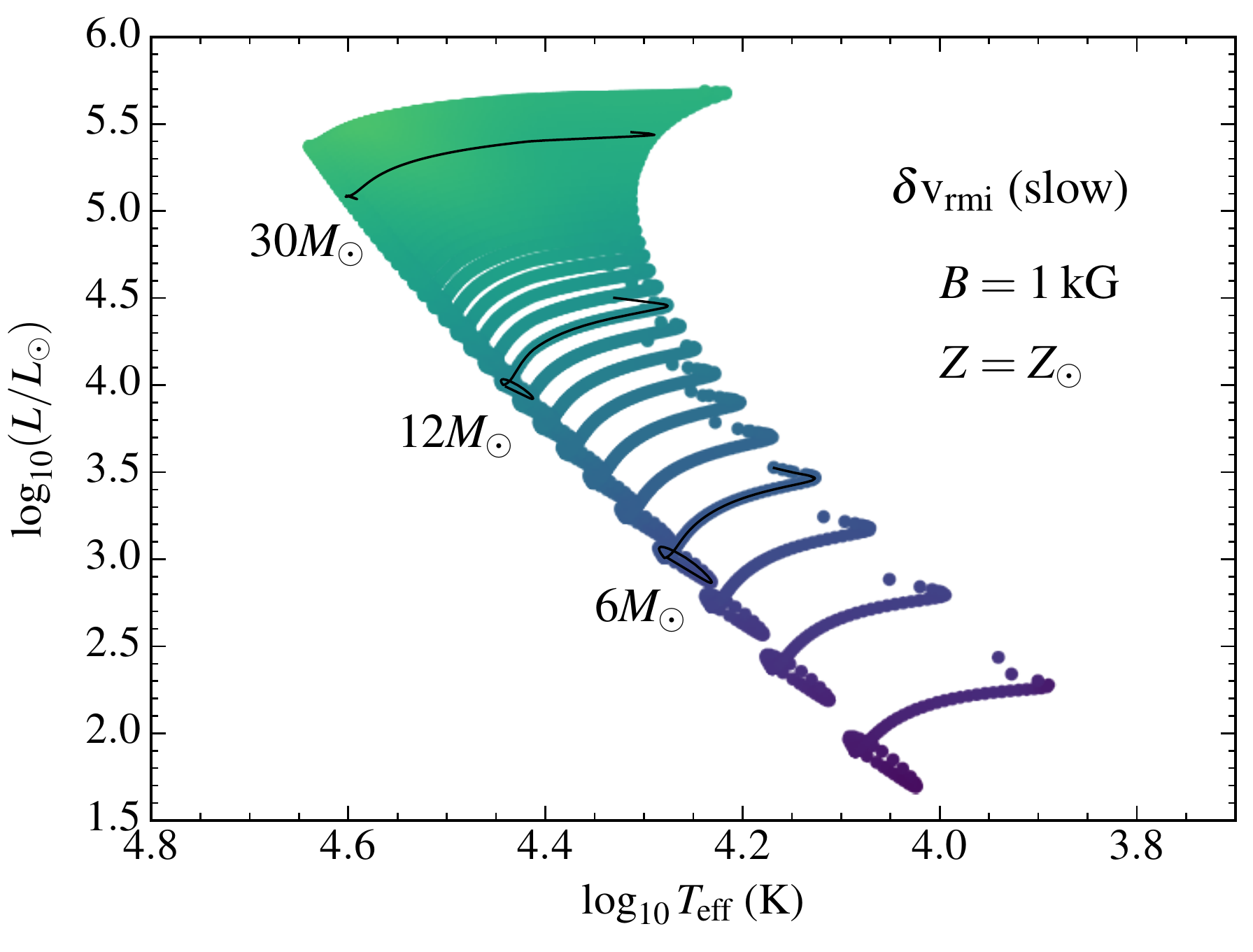}
\includegraphics*[width=1.055\columnwidth]{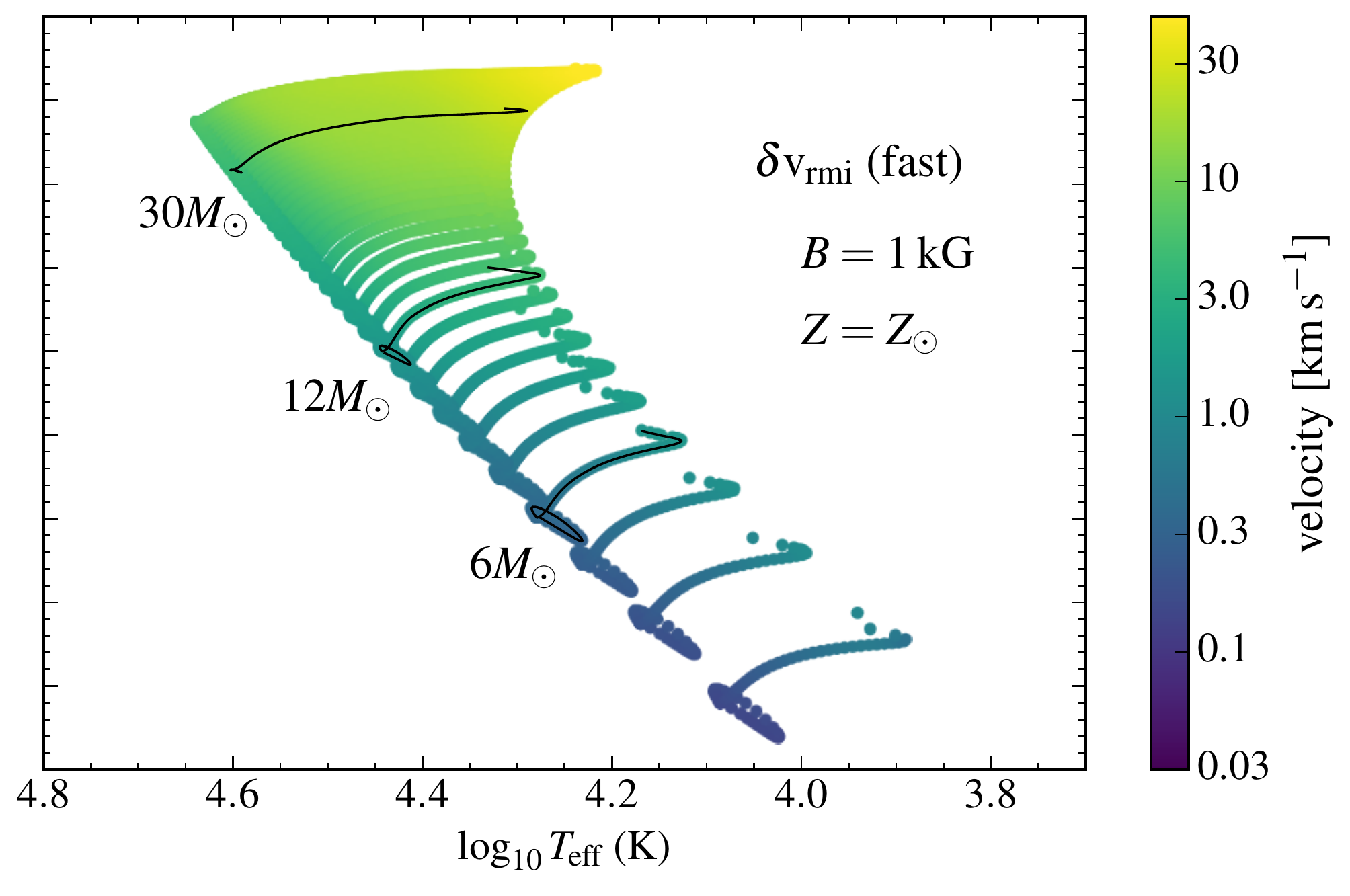}
\caption{RMI-driven velocity perturbation at $\tau = 10$ on the H-R diagram, as estimated from 
equations~(\ref{eq:drho_rmi})-(\ref{eq:dv_rmi}), for all of our models, and assuming 
a uniform magnetic field of $1$\,kG. The left panel shows the slow mode and the right 
panel shows the fast mode. The thin black lines show the trajectories followed in the 
H-R diagram by our three fiducial models, as labeled.}
\label{f:hr_dv}
\end{figure*}

Figure~\ref{f:zams} shows the magnitude of the RMI velocity perturbation at $\tau = 10$
obtained by evaluating equation~(\ref{eq:dv_rmi}) assuming a density amplitude as in equation~(\ref{eq:drho_rmi}).
By construction, the amplitude of the perturbation is a monotonically increasing function
of the ratio of radiation to gas pressure, which in turn is a monotonically increasing
function of ZAMS mass. Fluctuations are larger for fast modes than for slow modes by 
a factor of a few, exceeding $1$\,km\,s$^{-1}$ for $M\geq 7M_\odot$ and $M\geq 12M_\odot$
for fast and slow modes, respectively. At $M=40M_\odot$, the amplitudes are $6$\,km\,s$^{-1}$
$17$\,km\,s$^{-1}$ for slow and fast modes, respectively.

For comparison, we also compute the velocity fluctuation expected from gravity
waves excited by the iron convection zone. The kinetic energy density in internal 
gravity waves near the stellar surface is estimated by \citet{C09} as
\begin{equation}
\label{eq:energy_g-modes}
E_{\rm g}\sim \rho v_{g}^2\sim \rho_{\rm c} v_{\rm c}^2 \mathcal{M}_{\rm c},
\end{equation}
where $v_{\rm g}$ is the velocity associated with gravity waves near the
stellar surface, $\rho$ is the local density, and $\rho_{\rm c}$, $v_{\rm c}$ 
and $\mathcal{M}_{\rm c}$ are the density, velocity, and Mach number near the top 
of the iron convection zone. Equation~(\ref{eq:energy_g-modes}) assumes that the
kinetic energy densities in the line formation region and at the top of the
convection zones are comparable, because the volumes in these regions
are similar and because the wave energy flux is conserved. The expression
for the kinetic energy density of gravity waves at the top of the convection
zone \citep{goldreich_1990} is suitable for a discontinuous change in the Brunt-V\"ais\"al\"a
frequency $N^2$ at the convective-radiative transition; a more detailed analysis taking
into account the steep variation of this frequency with radius could yield a higher energy
conversion efficiency \citep{lecoanet_2013}. 
Equation~(\ref{eq:energy_g-modes}) results in a gravity wave velocity perturbation
\begin{equation}
\label{eq:vg_cantiello}
v_{\rm g}\sim v_{\rm c} \sqrt{\frac{\rho_{\rm c}}{\rho}\mathcal{M}_{\rm c}}.
\end{equation}
We compute $\rho_{\rm c}$, $v_{\rm c}$, and $\mathcal{M}_{\rm c}$ by
averaging over the top 1.5 pressure scale heights of the iron convection
zone, as done by \citet{C09}. We weight the average by the mass in each
zone in the radial range.

The bottom panel of Figure~\ref{f:zams} compares the estimated value of $v_{\rm g}$
from equation~(\ref{eq:vg_cantiello}) with the velocity perturbation introduced
by the RMI. Below $7M_\odot$, there is no iron convection zone operating in a significant
way and the resulting (very small) gravity wave velocities are not shown. Over the
range $7-40M_\odot$, the velocity fluctuations introduced by gravity waves
excited by convection zones are comparable to those introduced by the RMI.

Figure~\ref{f:zams} also shows the changes obtained in RMI-related quantities when varying
the magnetic field strength from $300$\,G to $3$\,kG (without any feedback
on the stellar structure). The ratio of magnetic to gas pressure varies
by an order of magnitude in all models. The radial range of the fast and
slow mode unstable regions change by a factor of order $\sim 10\%$, while
the unstable mass can vary by up to an order of magnitude. Note that the
variation of this regions is opposite for slow and fast modes: a stronger
magnetic field increases the region unstable to fast modes.

This inverse dependence on field strength also translates to the velocity 
perturbation. The fast mode velocity perturbation increases monotonically
with increasing magnetic field strength, while the slow mode velocity
does not increase for $v_{\rm A}> c_{\rm i}$. For a weaker field, the
slow mode velocity perturbation increases inversely with $(p_{\rm mag}/p_{\rm gas})^{1/2}$ 
due to the factor in square brackets in equation~(\ref{eq:dv_rmi}), which
is inversely proportional to the compressional energy in the mode (FS13).

Figure~\ref{f:hr_dv} shows velocity perturbations due to slow- and fast modes driven by the RMI
on an H-R diagram for all of our models, assuming a uniform magnetic
field of $1$\,kG. The magnitude of the slow
mode velocity perturbation does not vary significantly over the main sequence lifetime
of each star, depending most strongly on stellar mass. The fast
mode velocity perturbation increases with stellar age by up to about an order of magnitude
from its value at the ZAMS, in addition to depending on the mass of the star.

Figure~\ref{f:hr_dv} suggests that the RMI can readily account for the microturbulent
velocities discussed in \citet{C09} (c.f. their Figure 9), in particular for stars
below the luminosities (or metallicities) at which the iron convection zones disappear
or become greatly weakened. 

Variability timescales can be estimated by combining the value of the photospheric
pressure scale heights with the velocities apparent in Figure~\ref{f:hr_dv}. For
the $3M_\odot$ star, the photospheric scale height ranges from $10^{-3} - 10^{-2}R_\odot$
from ZAMS to the end of the main sequence. RMI-induced velocities of $0.1-1$\,km\,s$^{-1}$,
imply variability timescales ranging from tens of minutes to days. For our $40M_\odot$
star, both the scale heights and the velocity range are factor of about $10$ larger, 
so the variability timescales should be similar. While the fastest growing mode
of the RMI has wavelengths comparable to the gas pressure scale height, smaller wavelengths
are also destabilized (BS03), with even shorter variability timescales being possible (FS13).

\section{Summary and Discussion}
\label{sec:summary}

We have studied the conditions for the generation sub-photospheric fluctuations
by unstable magnetosonic waves in the radiative envelopes of intermediate-
to high mass stars with a magnetic field. A grid of solar metallicity models 
covering the mass range $3-40M_\odot$ was used to evaluate the instability conditions for
the Radiation-Driven Magneto-Acoustic Instability (RMI, also known as
the photon bubble instability), in which slow- and fast magnetosonic
modes are destabilized in optically-thick media by a background radiative 
flux when a magnetic field is present. Our main results are the following:
\newline

\noindent
1. For a uniform magnetic field of $\sim 1$\,kG strength, the RMI
   operates in all models we studied. Fast magnetosonic modes are
   unstable to a depth of a few pressure scale heights below the photosphere,
   while slow modes are unstable to depths beyond the iron convection
   zone (Figures~\ref{f:profiles} and \ref{f:evolution}). 
   \newline

\noindent
2. Driving due to the radiative flux acting on opacity variations never
   dominates over the RMI for slow modes at short wavelengths, regardless of magnetic field
   strength or ratio of radiation to gas pressure, so long as $\Theta_\rho < 1$,
   as is the case in all of our models. For fast modes, driving due to opacity variations dominates for 
   field strengths $\lesssim 300$\,G (Figure~\ref{f:bmin}, equation~\ref{eq:bmin}) 
   or beyond the depth at which RMI driving of fast modes is stabilized by radiative
   diffusion (Figure~\ref{f:profiles}).
   \newline

\noindent
3. Assuming that the magnetic field is generated at the iron convection
   zone results in fast modes being driven by the radiative flux acting
   on opacity variations (Figure~\ref{f:profiles_flux-freeze}). Slow modes 
   are still driven by the RMI to similar depths as with larger field strengths,
   but with a smaller growth rate given the weak forcing when $v_{\rm A}\ll c_{\rm i}$
   (equation~\ref{eq:slow_instability_criterion}).
   \newline 
  
\noindent
4. The implied magnitude of velocity fluctuations for slow- and fast modes
   in the case of a $\sim 1$\,kG spans from $\sim 0.1 - 10$\,km\,s$^{-1}$,
   with the amplitudes being a monotonically increasing function of the
   ratio of radiation to gas pressure, or alternatively, of the stellar
   mass (Figure~\ref{f:zams} and \ref{f:hr_dv}). These amplitudes are comparable
   to those expected from the excitation of gravity waves at the top of
   the iron convection zone, whenever the latter is present.
   When considering the RMI-driven velocities and the photospheric
   pressure scale height, the implied variability timescales range
   from minutes to days (\S\ref{s:mass}).
   \newline

\noindent
5. Changing the magnetic field strength by a factor of a few does not
   alter our results qualitatively (Figure~\ref{f:zams}). The velocity
   amplitude of slow modes is a monotonically increasing function of the
   ratio of magnetic to gas pressure, while the amplitude of slow modes 
   increases for smaller values of $p_{\rm mag}/p_{\rm gas}$.
   \newline

\noindent
6. Aside from significantly modifying sub-surface convection zones, lowering the metallicity
   results in minor quantitative modifications in the properties of the
   RMI, for a fixed background magnetic field (Figure~\ref{f:metallicity}).
   \newline

The instability criteria studied in this paper are local, hence we cannot say
anything about its effect on spatial scales larger than the gas pressure scale height. 
In particular, kinetic energy generated via the RMI can couple to global stellar 
oscillation modes, resulting in excitation of specific frequencies or in transfer 
of energy to other spatial scales. To assess these effects, global radiation-magnetohydrodynamic 
simulations of stellar envelopes are required.

Our velocity estimates are rough and subject to significant revision. In particular,
the saturation amplitude of the RMI has never been studied in three spatial dimensions
in the regime in which $p_{\rm rad}, p_{\rm mag}\lesssim p_{\rm gas}$, as is applicable
here. Use of equation~(\ref{eq:drho_rmi}) is therefore the best estimate we can make at
the moment. Future work should address the saturation amplitude of this instability
in all regimes of interest, using all spatial dimensions, and including realistic
stellar opacities. 

Similar uncertainties apply to the transition to the optically thin regime. At
low enough optical depth, the conditions for the RMI to operate are
no longer valid. Nevertheless, radiation can still impart significant momentum 
to matter -- if only at specific wavelengths -- thus interesting phenomena are
likely to occur in the presence of a strong magnetic field. The development
of suitable tools to study this problem will provide insight on this question.

\section*{Acknowledgements}
We thank the referee, Steven Shore, for constructive comments that improved the manuscript.
We also thank Huib Henrichs for stimulating discussions. 
KS was supported in part by the University of Alberta Research
Experience (UARE) program and by the Inspire Fellowship Programme of
the Indian Department of Science and Technology. RF acknowledges
support from Natural Sciences and Engineering Research Council (NSERC) 
of Canada and from the Faculty of Science at
the University of Alberta. This research was enabled in part by
support provided by WestGrid (www.westgrid.ca) and Compute Canada
(www.computecanada.ca). Computations were performed at the \emph{Jasper} and \emph{Orcinus} 
compute clusters.
This research also used storage resources (repository 2058) of the National Energy Research
Scientific Computing Center (NERSC), which is supported by the Office of
Science of the U.S. Department of Energy under Contract No. DE-AC02-05CH11231.

\appendix


\bibliographystyle{mn2e}
\bibliography{ms}

\label{lastpage}

\end{document}